# Contactless Thermal Characterization of High Temperature Test Chamber

Z. Szűcs[1], Gy. Bognár[1], V. Székely[1], M. Rencz[1,2]

[1] Budapest University of Technology and Economics, Hungary, <szucs|bognar|szekely|rencz >@eet.bme.hu
[2] MicReD, Budapest, Hungary, rencz@micred.com

*Abstract* – In this paper the methodology and the results of a contactless thermal characterization of a high temperature test chamber will be introduced. The test chamber is used for fatigue testing of different MEMS devices where the homogenous temperature distribution within the close proximity from the heating filaments is very important. Our aim was to characterize the evolving temperature distribution inside the test chamber. In order to achieve smaller time constant a new contactless sensor card was developed. The contactless thermal characterization method introduced in this paper enables in situ heat distribution measurement inside the test chamber during operation, with the detection of potentially uneven heat distribution.

## I. INTRODUCTION

The elevated temperature encountered in different packaged electronic devices demands the application of careful temperature-aware design methodologies and electro-thermal simulations of the entire system.

The results of various electro-thermal simulations and models in most of the cases yield good approximations. These models take the coupled effects of the surroundings and other dissipating elements in the system into consideration. However, the simulation time may take a long time, and after any changes in the system or in any parameters of the surroundings, the simulation should be executed again.

Our goal was to develop a tool with which the actual temperature distribution of an electronic system can be characterized (e.g. a board in an equipment may be checked instantaneously). Using a contactless temperature measurement procedure the heat distribution and the places of high dissipation elements on an operating PWB (Printed Wiring Board) (e.g.: PCI or AGP cards) can be measured and localized in a dense rack system, where only a thin measuring board could be inserted between the operating cards.

In other applications, the uniform heat distribution may be the most important issue during operation. This is the case e.g. at a high temperature test chamber, where the evolving homogenous temperature distribution – within the close proximity of the heating filaments – is elementary. This is needed to maintain that during accelerated fatigue testing every part of the device under test (e.g.: packaged integrated circuits, MEMS systems, assembled PWB boards) is heated up or cooled down to the same temperature.

In the case of a test chamber, mentioned before, only a contactless temperature measuring method can be used in order to measure the evolving heat distribution. The main reason for this is the small slot of the device where the testing device may be introduced.

In our expectation, by using a contactless temperature measurement procedure the possible inequality of the heat distribution and the places of the high dissipation elements or different hot-spots in an operating high temperature test chamber can be measured and localized.

There are several contactless temperature measurement methods, but the price and usability determine their application area [2]. The different types of thermographic cameras are not only expensive but it is impossible to insert them for example between the cards of a system or into a test chamber with thin slot. That is why the heat distribution can not be visualized in this way. An additional problem is that cooled type IR cameras need liquid gases during the operation [3].

Uncooled IR detectors realized by MEMS technology are mostly based on pyroelectric materials, microbolometer or thermopile technology [4]. Thanks to the MEMS technology small sensor array can be realized as well on a silicon die similar to CCD sensors. The only difference is in the sensing method. The CCD sensors are sensitive for the near infrared radiation as well (upto 900 nm). This range of spectra is used mainly for night vision application. To determine the temperature of a distant object the sensing of the far infrared radiation is needed (some μm range) if the temperature of the devices is between 0 and 100°C. The main advantage of these sensors is the relatively small time constant (approx. 10 ms). Further disadvantage is the additional IR optical elements which are needed to get a relatively wide viewing angle. With these additional elements however, the whole IR sensor system may not be placed inside the test chamber.

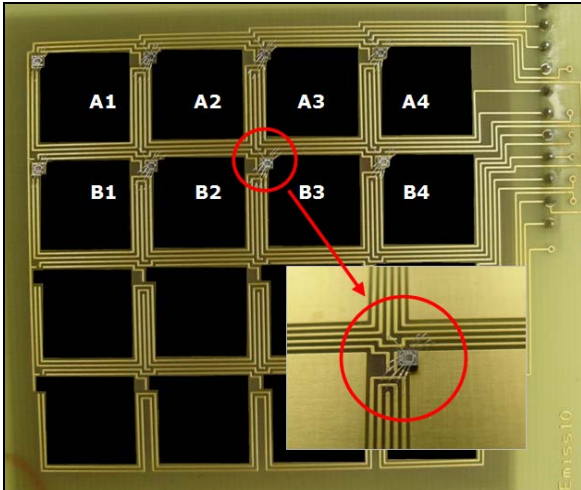

Figure 1. – The temperature sensor dies attached onto the pixels of the sensor card

In this paper a novel IR measurement method is introduced to characterize the high temperature test chamber.

## II. THE DEVELOPMENT OF THE SENSOR CARD

In our conception a contact-less method was used to localize different thermal hot-spots. This method can be applicable for sensing the temperature distribution map of e.g. a card in a dense rack system or a heat chamber, whereto only a thin board can be fit.

The IR sensor card [5] was realized on a normal (1.55 mm) printed wiring board with a special metallization pattern. In previous articles [7][8] the applicability of this sensor card was demonstrated. The main problems were the high thermal time constant and the low resolution. That device was able to detect the hotspots of a PCI card in an operating computer rack, but a higher resolution was required in that application.

In the first experiment a 4 x 4 matrix was created (Figure 1). The function of each square-shaped 10 mm x 10 mm metal "pixel" on the card is to absorb of IR radiation. The temperature rise due to the inbound energy is sensed by using bare thermal test-chips, attached directly onto the copper plate of the "pixel" to minimize the thermal resistance between the central point of the pixel and the thermal test-chip.

The temperature sensing is based on the changes of the integrated diode's forward voltage. The temperature of the chip related to the temperature of the pixel. There is a thermal resistance between the plate of the pixel and the active area of the chip, the value of which depends on the type and the quality of the die attach. The sensed forward voltage variation is digitalized by using a custom A/D converter. The applied thermal test chips provide an output frequency depending on the temperature rise of the "pixel". The design of the test chip used in this application is derived from the TMC family of TIMA and MicReD [6].

In order to minimize the time constant of each pixel the thermal capacitance had to be decreased by selecting a board (1.55 mm) and metallization layer (35 μm) as thin as possible.

To facilitate good quality ultrasonic bonding, the copper surface on the upper side of the measurement card was plated with a 2 μm gold layer. The entire board was painted black to achieve a better absorption of the radiated heat.

Unfortunately the thermal time constant of the sensor card was too high caused by the low thermal conductance of the die-attach which was realized by using an electrically non-conductive adhesive. Another problem was the low resolution.

That is why a new sensor card was needed to develop (Figure 2). The temperature sensor chips were substituted by simple, minimal area surface mounted diodes (SMD) (0.25mm×0.50mm). The forward voltage of the diodes depends on the absolute temperature of the junction, which is related to the absorbed IR radiation of the black-colored metal plates.

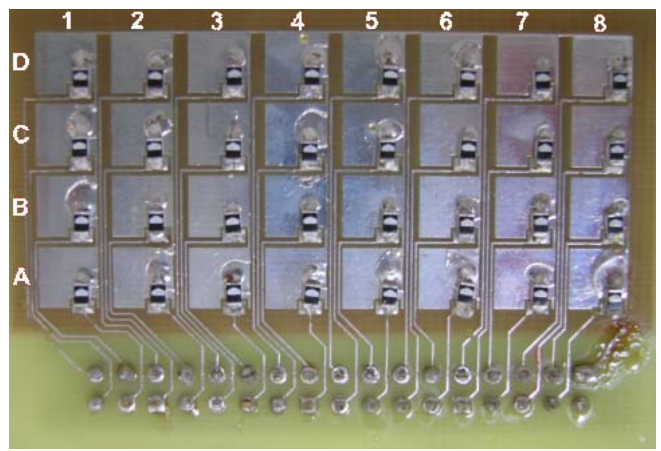

Figure 2. – The new 4x8 sensor card before being painted black

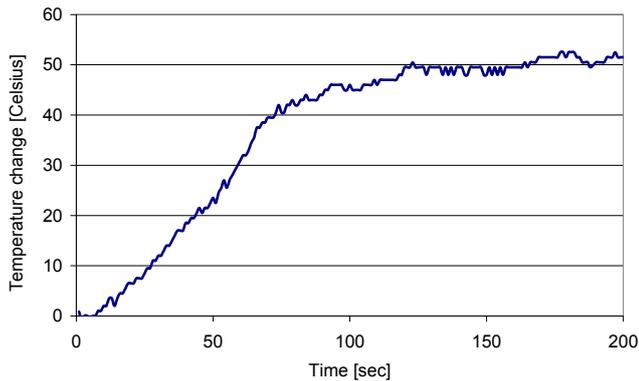

Figure 3. – The thermal transient of the B2 pixel

2 mV change sensed by the diodes means 1 °C temperature change. Each diode of the sensor matrix can be selected by using an analog multiplexer circuit. To convert these sensed analog voltages to digital values, processable by the microcontroller, an AD623 amplifier and the 10 bit A/D converter of the microcontroller were used. A simple read-out circuitry controlled by Atmel microcontroller was built to forward the acquired values to a PC. The communication between the PC and the circuit is realized by serial link (RS-232) communication as in the case of the previous sensor card.

In order to reduce the thermal time constant and achieve finer lateral resolution the size of the pixels was changed. On the new sensor card an 8 × 4 matrix was created by using 5 mm × 5 mm square pixels. The sensor card was realized on a printed wired board with the same thickness (1.55 mm) used before. The thickness of the copper plate was 35 µm. In order to minimize the thermal resistance between the pixel and the junction of the diode we paid attention to contact the substrate of the diodes to the metal plates, as the higher heat conduction area results in lower thermal resistance. The substrate of the applied SMD diodes was the cathode contact. This contact was placed on to the metal pixel and the diodes were attached to the surface by using reflow soldering method. The thermal resistance between the diodes and the copper plates is lower than in the case of the first board. It was caused by the greater interface area and higher thermal conductance between the substrate of the diode and the pixel.

Only a 1.25 mm gap was kept for wiring between the pixels. All cathodes of the diodes were connected to ground potential. Only 178 µm wide copper wires connect the adjacent pixels. This results in very low cross heat flow between the pixels. The entire board was painted black to facilitate better absorption of the radiated heat.

The thermal time constant of the new sensor card was decreased compared to the previous sensor card which is proven by the calibration measurements.

*In the frame of calibration procedure a heated black body was placed in front of the sensor card and the thermal transient of the pixels was monitored.* The thermal transient of the B2 pixel can be seen in the Figure 3. It can be recognized that the system approaches equilibrium state after 120 s, which is lower than the previously measured 240 s. This value unfortunately means that very fast temperature changes can not be followed by using this method. Of course by using such sensor card, realized on PWB, we can not exploit the 10 ms time-constant of the integrated pyroelectric type sensors.

III. THE MEASUREMENT SETUP

A small, computer-controlled test chamber was designed and created at BME – Department of Electron Devices, for the purpose of testing micro-electromechanical and other microsystems. The design of the interior is intentionally flat to obtain the advantage of fast temperature response over the big volume, commercial test chambers.

Temperature Cycle Testing (TCT) or simply temperature cycling, determines the ability of parts to resist extremely low and extremely high temperatures, as well as their ability to withstand cyclical exposures to these temperatures. A mechanical failure resulting from cyclic thermo-mechanical load is known as a fatigue failure, so temperature cycling primarily accelerates fatigue failures. Failure mechanisms accelerated by temperature cycling include die cracking, package cracking, neck/heel/wire breaks, and bond lifting

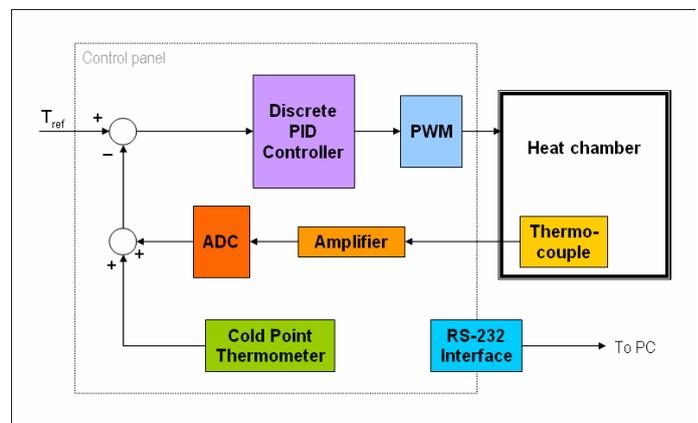

Figure 4. – Block diagram of thermal test chamber

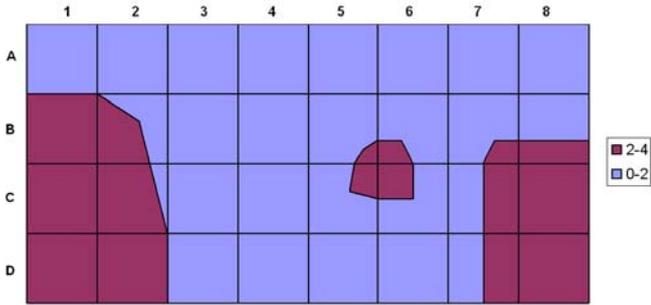

Figure 7. – The evolving temperature distribution after 2 minutes inside the chamber *(numbers show temperature difference against the measured 81.5 °C average temperature sensed by the thermocouple)*

In this tester equipment the uniform heat distribution is a very important requirement. Testing the chamber is a good example where our testing method is the best measurement approach.

The device is able to produce temperature values in the range of 20…250 °C (±1 °C) with the help of a digital PID controller. The temperature is sensed by a thermocouple and the heating of the 200 cm$^3$ chamber is PWM modulated (Figure 4).

During the measurement the black-colored sensor card was placed into the high temperature test chamber, in parallel with the heating filament. We supposed that the nature of heat distribution does not depend on the operational temperature, which can be set by programming the PID controller. The test chamber was operated at 100 °C. The diodes on the sensor card were calibrated between 20 and 120 °C before the measurements.

The sensor card was attached to the door of the chamber in order to hold the card accurately in the middle of the slot, in 1 cm distance from the heated wires. The *B*, *C* and *D* rows of pixels were placed under the heating filaments. Above the *A* row, there was no filaments and this row was in the proximity to the door.

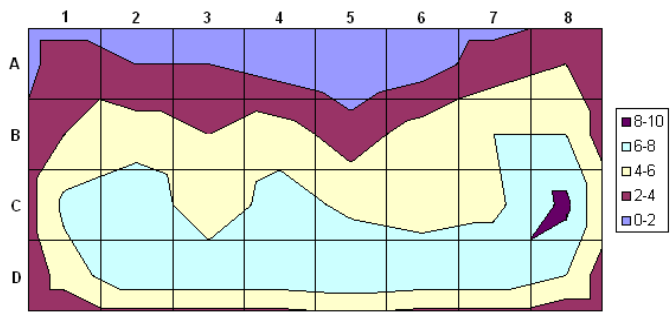

Figure 5. – The measured temperature distribution after 10 minutes inside the chamber *(numbers show temperature difference against the measured average temperature sensed by the thermocouple)*

## IV. MEASUREMENT RESULTS

The measured temperature elevation values obtained by the sensor card after 120 s, proved that the heat distribution inside the chamber is almost uniform. Only 2 °C difference can be visualized in Figure 7. near to the door of the chamber.

The steady state results in Figure 5. – after 600 s – show, that near to the door uneven distribution was detected. It was caused by the inadequate thermal isolation between the interior of the chamber and the ambient.

The measurement proved that the exact place of the heating filament can be localized and the heat distribution can be measured.

Under the heating filaments the evolving temperature distribution was homogenous and the results proved the applicability of the test chamber for TCT accelerated fatigue testing of different MEMS or assembled PWB devices.

## V. SIMULATION

Simulations have been carried out for cross-verification purposes and to demonstrate the applicability of the sensor card inside the test chamber. A simple model of the high temperature test chamber was created. A steady-state solution of our measuring system in the test chamber considering the heat radiation was done by FLOTHERM 7.2 program [10].

The steady state results show that inside the chamber the evolving heat distribution in 1 cm distance from the filaments is uniform. In Figure 6 the black parallel lines symbolize the heating filaments and the evolving temperature distribution along the sensor card can be seen. This result shows good correspondence to the results from the measurement.

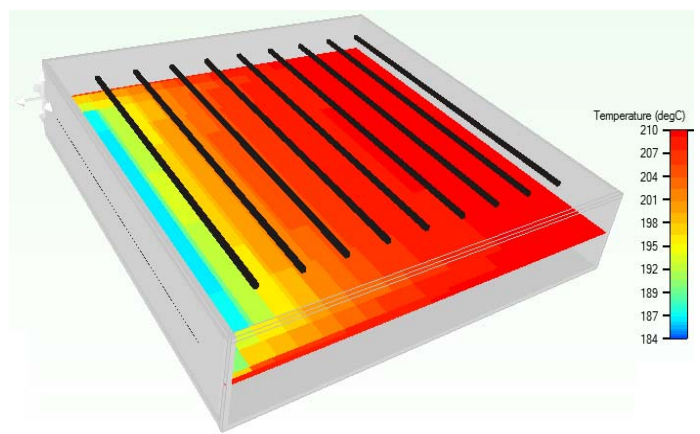

Figure 6. – The steady state simulation results

The simulation proved the feasibility of determining the heat distribution inside the high temperature test chamber.

## VI. SUMMARY

We have presented a contactless method to localize thermal hot-spots and lateral heat distribution inside a high temperature test chamber for testing different microelectronic and MEMS devices.

We have succeeded to decrease the thermal time constant of the sensor card by using minimal area SMD diodes and realizing a low thermal resistance heat path between the pixel and the junction. For further decreasing the time constant a thinner PWB board, maybe back-drilled one, under the metal plates could be applied.

We have successfully proved that the heat distribution inside the high temperature test chamber, under the filaments is uniform. So this equipment is suitable for testing and thermal measurement of integrated circuits, MEMS devices or other micro-sized samples.

## VII. ACKNOWLEDGEMENT

This work was supported by the PATENT DfMM EU 507255 and OTKA TUDISK TS049893 projects. The authors would like to thank for the help of the Research Institute for Technical Physics and Materials Science of Hungary (MFA).